\documentclass[aps, prl, amsmath, amssymb, amsfonts, twocolumn, showpacs, floatfix]{revtex4}
\usepackage{graphicx}
\usepackage{subfigure}
\usepackage{dcolumn}
\usepackage{bm}
\usepackage{pst-node}
\usepackage{epsfig}

\newcommand     {\beq}[1]         { \begin{equation} #1 \end{equation} }
\newcommand     {\beqa}[1]        { \begin{eqnarray} #1 \end{eqnarray} }

\newcommand     {\ep}             { \varepsilon }

\begin{document}

\title{Fiber bundle model with stick-slip dynamics}

\author{Zolt\'an Hal\'asz and Ferenc Kun$\footnote{Electronic address:feri@dtp.atomki.hu}$}
\affiliation{
Department of Theoretical Physics, University of Debrecen, P.\
O.\ Box:5, H-4010 Debrecen, Hungary}
\date{\today}
\begin{abstract}
We propose a generic model to describe the mechanical response and
failure of systems which undergo a series of stick-slip events when
subjected to an external load. We model the system as a bundle of fibers,
where single fibers can gradually increase their relaxed length 
with a stick-slip mechanism activated by the increasing load. 
We determine the constitutive equation of the system and show by
analytical calculations that on the macro-scale a plastic response
emerges followed by a hardening or softening regime. Releasing the
load, an irreversible permanent deformation occurs which depends on
the properties of sliding events.  For quenched and annealed disorder
of the failure thresholds the same qualitative behavior is found,
however, in the annealed case the plastic regime is more pronounced.
\end{abstract}  
\maketitle

There is a large variety of systems which undergo conformational
changes when subjected to external mechanical loads. 
The dynamics of spatial rearrangements can usually be described by
the stick-slip mechanism, i.e.\ when the local load exceeds some threshold
value, subunits of the system increase their length leading to
relaxation. The stored length which can be activated by external
loading, typically arises due to the existence of frictional contacts 
or chemical bonding between subunits. 
Several examples of such systems can be mentioned from the molecular
scale organization of spider silk
\cite{spider_vollrath_softmatter2006}, through the chains of magnetic 
beads in magneto-rheological fluids on the meso-scale
\cite{furst_gast_prl1999,biswal_gast_pre2003}, to the wire nets used
to protect roads from 
rockfalls in mountains \cite{rockfall_1} on the
macro-scale. Experiments have revealed that the  
surprisingly high fracture toughness of spider silk is partly caused
by the presence of blobs of protein molecules with a folded hairpin
structure, which get unfolded under
external loading \cite{spider_vollrath_softmatter2006}. A similar
mechanical response has also been observed for 
biological tissues composed of interconnected bundles of fibrils
\cite{fiber_bio_2007}.  
In magneto-rheological fluids particles of permanent magnetic moment
aggregate and form chains aligned with the external field, which then
modify the rheological properties of the fluid. Stretching and bending
chains of particles of micrometer size by optical tweezers revealed
that the anisotropic magnetic interaction and the frictional contact
of particles result in a noisy response, i.e.\ particles undergo
subsequent rearrangements increasing the length of the chain and
reducing the reactant force
\cite{furst_gast_prl1999,biswal_gast_pre2003}. 
A similar mechanism is exploited on the macro-scale in wire nets,
which cover steep walls in mountains protecting roads from rockfalls. 
In order to dissipate the kinetic energy of falling boulders,
the wires form rings which can slide and collapse without breaking
resulting in a large dissipation but keeping the integrity of the net
\cite{rockfall_1}. 

In the present paper we introduce a micro-mechanical model which
captures the main ingredients of the mechanical response of system
which undergo conformational changes with stick-slip mechanism. Our  
model construction is based on fiber bundle models
\cite{hansen_crossover_prl,tommasi_prl_2008,kun_epjb_17_269_2000,kim_prl_94_025501_2005,naoki_prl2008,raul_varint_2002} extended in such a 
way that fibers undergo sliding events which gradually increase their
relaxed length when the local load exceeds some threshold values. After
a large number of sliding 
events, fibers may also fail under a large enough external load. We
derive the constitutive equation of the system and show analytically
that the 
sliding mechanism leads to macroscopic plasticity of the bundle and
permanent deformation remains after the load has been released. 
We explore the case of both quenched and annealed disorder of sliding
thresholds. 

\begin{figure}
  \begin{center}
\epsfig{bbllx=0,bblly=0,bburx=280,bbury=350,file=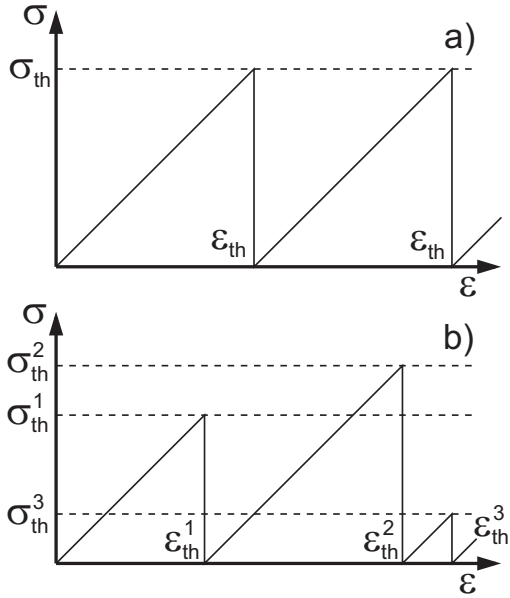,width=6.5cm}
   \caption{Response of a single fiber for quenched $(a)$ and annealed
$(b)$ disorder. When the failure threshold is 
reached the load on the fiber drops down to zero. The threshold can be
constant $(a)$, or new ones can be drawn from the
same distribution $(b)$.}
\label{fig:single_fiber}
  \end{center}
\end{figure}

Our model consists of $N$ parallel fibers which have identical 
elastic properties characterized by the Young modulus $E$
\cite{hansen_crossover_prl,tommasi_prl_2008,kun_epjb_17_269_2000,kim_prl_94_025501_2005,naoki_prl2008,raul_varint_2002}.
Under an increasing external load $\sigma$ parallel to the fibers'
direction, the fibers exhibit a linearly elastic behavior until the
local deformation $\ep_i$ reaches a threshold value
$\ep_{th}^i$, $i=1,\ldots , N$. The novel element of the model is
that at the threshold the fiber does not break, instead it suffers
sliding, i.e.\ its relaxed length increases until the fiber becomes
capable to sustain the remaining load. The sliding threshold of fibers
$\ep_{th}$ is a random 
variable with a probability density $p(\ep_{th})$ and a
distribution function $P(\ep_{th})$ defined over the domain
$[\ep_{th}^{min},\ep_{th}^{max}]$. The sliding event is 
instantaneous, i.e.\ it does not take time. After the relaxation the
fiber can be loaded again which may lead to further slidings. 
The total number of allowed sliding events (steps of extension)
$k_{max}$ is an important 
parameter of the model. When the fibers can slide more than ones
$k_{max}>1$, they either keep the initial sliding threshold for all
slidings (quenched disorder), or can obtain a new threshold value each
time from the same probability distribution (annealed disorder).
The constitutive behavior of single fibers is illustrated by Fig.\
\ref{fig:single_fiber} for both types of disorder. 

The sliding of a fiber means that its equilibrium length gets suddenly
increased, which has the consequence that the fiber can only keep a
lower load $\sigma_i = E(\ep-\ep_{th}^i)$,
where $\sigma_i$ denotes the local load of fiber $i$, $\ep_{th}^i$ is
its sliding thresholds and $\ep$ is the macroscopic strain of the
system.  
Under a fixed external load $\sigma$, the other
fibers of the bundle have to overtake the load dropped by the
one which has just been extended. For the 
load redistribution we assume an infinite range of interaction, i.e.\
global load sharing
\cite{raul_varint_2002,kun_epjb_17_269_2000}. However, it does not
imply that the load is everywhere the same in the bundle. At a 
given macroscopic deformation $\ep$ the fibers have suffered a
different number $k_i$ of slidings occurring at different threshold
values, hence the load they keep will be different:
$\sigma_i=E(\ep-k_i\ep_{th}^i)$ for quenched, and
$\sigma_i=E(\ep-\ep_{th}^{i,1}-\ep_{th}^{i,2}-\cdots -\ep_{th}^{i,k_i})$
for annealed disorder ($i=1, \ldots, N$), respectively. (A fuse model
with a slightly similar microscopic dynamics was considered in Ref.\
\cite{sornette_fusefatigue_prl1992}, where fuses burn out due to gradual
overheating.)

We start the analysis with the case of quenched disorder and assume
that only a single sliding is allowed for the fibers
$k_{max}=1$. At the 
macroscopic deformation $\ep$ the fibers with sliding 
thresholds $\ep_{th}^i \geq \ep$ are still intact and have the local
load $E\ep$. However, those fibers which have already suffered a
sliding $\ep_{th}^i < \ep$ keep only the load 
$E(\ep-\ep_{th}^i)$ so that the macroscopic constitutive equation 
follows as
\beq{
\sigma(\varepsilon) = E\ep\left[1-P(E\ep) \right] +
\int\limits_{\ep_{th}^{min}}^{\ep}p(E\ep_1)E\left(\ep-\ep_1\right)d\ep_1. 
\label{eq:constit_1}}
Note that the integral in the second term is performed over the 
entire loading history of the system. In the integrand the probability
density $p$ of the thresholds occurs since the load of the extended fibers 
depends on the precise value of the deformation where the sliding initiated.
Allowing for two sliding events $k_{max}=2$, at the macroscopic
deformation $\ep$ those fibers slided exactly once whose sliding
threshold falls in the range $\ep/2 < \ep_{th}^i < \ep$, while the
ones with $\ep_{th}^i < \ep/2$ already suffered two slidings, hence, 
$\sigma (\ep)$ can be written as 
\beqa{
\sigma(\varepsilon) &=& E\ep\left[1-P(E\ep) \right] + \nonumber
\int_{\ep/2}^{\ep}p(E\ep_1)E\left(\ep-\ep_1\right)d\ep_1 \\ 
&+&\int_{\ep_{th}^{min}}^{\ep/2}p(E\ep_1)E\left(\ep-2\ep_1\right)d\ep_1.
\label{eq:constit_2}}
\begin{figure}
  \begin{center}
  \epsfig{bbllx=35,bblly=515,bburx=325,bbury=755,file=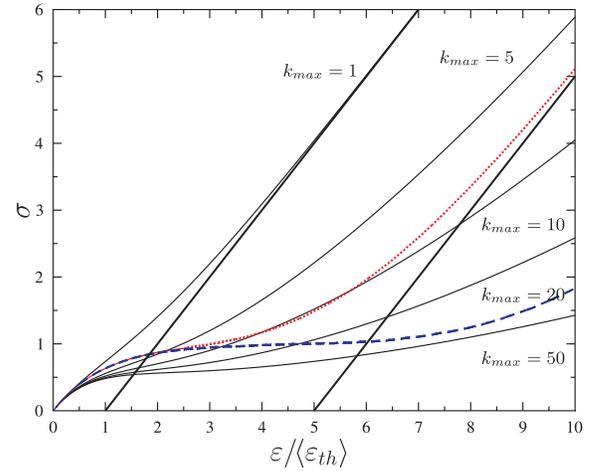,width=7.8cm}
  \caption{({\it Color online}) Constitutive behavior of the bundle with exponentially
distributed quenched failure thresholds. $\sigma(\ep)$ tends to an
asymptotic linear response preceded by a more and more horizontal
plateau when $k_{max}$ increases. The intersection of the asymptotic
straight lines with the horizontal axis indicates the value of
$\ep_r^{max}$ obtained when unloading the system. For annealed
disorder the cases of $k_{max}=5$ (red, dotted line) and 10 (blue,
dashed line) are shown.
}
  \label{fig:const_exp}
  \end{center}
\end{figure}

For any arbitrary value of $k_{max}$ the sliding threshold of fibers
which are intact or suffered a $k$ number of sliding events $1 \leq k
\leq k_{max}$ fall in the following sub-intervals of
$\left[\ep_{th}^{min},\ep_{th}^{max} \right]$ 
\beqa{
 \ep &<& \ep_{th}^i, \ \ \ \ \ \ \ \ \ \ \ \mbox{intact}, \nonumber \\
 \frac{\ep}{k+1} &<&  \ep_{th}^i < \frac{\ep}{k} \ \ \ \ \ \
\mbox{sliding k times, 
where} \ \ k < k_{max} \nonumber \\
 0 &<&  \ep_{th}^i < \frac{\ep}{k_{max}} \ \mbox{sliding} \ \
k_{max} \ \ \mbox{times}.
\label{eq:subinterv}}
The general form of the macroscopic constitutive equation can be
obtained by integrating the load kept by the above subsets of fibers
\beqa{
&&  \sigma(\ep) =  E\ep\left[1-P(E\ep) \right]  \nonumber \\ 
&&+ \displaystyle{\sum_{k=1}^{k_{max}-1}\nonumber
\int_{\ep/(k+1)}^{\ep/k}p(E\ep_1)E\left(\ep-k\ep_1\right)d\ep_1}
\nonumber \\ 
&&+  \displaystyle{\int_{\ep_{th}^{min}}^{\ep/k_{max}}p(E\ep_1) 
E\left(\ep-k_{max}\ep_1\right)d\ep_1}.   
\label{eq:bundle_constit}}
\begin{figure}
  \begin{center}
\epsfig{bbllx=160,bblly=393,bburx=478,bbury=651,file=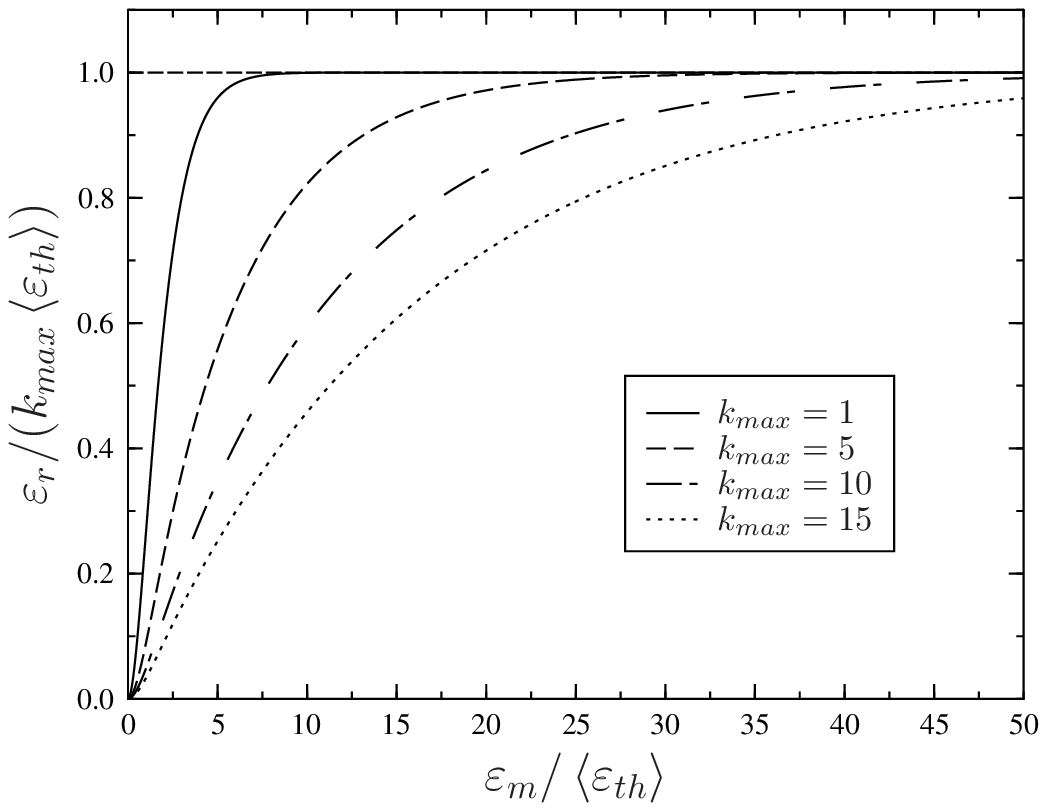,width=8.0cm}
   \caption{Remaining deformation $\ep_r$ as a function of the maximum
deformation $\ep_m$ reached before the unloading starts for different
values of $k_{max}$. $\ep_r$ and $\ep_m$ are normalized by the average
sliding threshold $\left<\ep_{th}\right>$ and the maximum value
$\ep_r^{max}$, respectively.
}
\label{fig:remaining}
  \end{center}
\end{figure}

Note that in Eq.\ (\ref{eq:bundle_constit}) the fibers retain their
initial stiffness after suffering $k_{max}$ sliding events, which is
expressed by the last term. It can be seen that for small
deformations $\ep \to 0$ only the intact fibers are relevant which
result in a macroscopic linear behavior with a Young modulus equal to
that of the individual fibers. 
For very large deformations $\ep \to
\infty$, practically all fibers have suffered $k_{max}$ restructuring
events so that only the last term of Eq.\ (\ref{eq:bundle_constit})
survives which can be further written as 
\beq{
\sigma(\ep) \sim E\ep -k_{max}E \int_{\ep_{th}^{min}}^{\ep/k_{max}}
p(\ep_1)\ep_1 d\ep_1. 
}
The integral in the second term converges to the average value of the
thresholds $\left<\ep_{th} \right>$, hence, we obtain $\sigma(\ep)
\sim E\ep - k_{max}E\left<\ep_{th} \right>$. 
The result implies that the bundle has an asymptotic linear behavior with the
initial value of the Young modulus. When unloading the
system $\sigma(\ep) \to 0$, the fibers simply relax with a linearly
elastic response since there is not any mechanism in the model to
decrease the equilibrium length of fibers. It has the consequence that
as the load is released, the system relaxes along a straight line of
slope $E$ equal to the Young modulus of fibers and 
an irreversible remaining deformation occurs $\ep_r$, whose maximum
value $\ep_r^{max}$ is proportional to the average slip length
$\left<\ep_{th}\right>$ and the number of sliding events $k_{max}$
allowed $\ep_r^{max} = k_{max}\left<\ep_{th}\right>$.
Unloading the bundle after
reaching a macroscopic deformation $\ep_{m}$ the remaining irreversible
deformation $\ep_r$ can be obtained as $\ep_r = \ep_{m} - \sigma(\ep_m)/E$.
For the purpose of explicit calculations we used a Weibull
distribution for the sliding thresholds with the cumulative
distribution $P(\ep_{th}) = 1-e^{-(\ep_{th}/\lambda)^m}$ over the
range $0\leq \ep_{th}<+\infty$,
where $\lambda$ sets the scale of the threshold values and $m$ denotes
the Weibull exponent. In all the calculations $\lambda=1$ was chosen.
The constitutive curve of the model is presented in Fig.\
\ref{fig:const_exp} for the case of $m=1$, i.e.\ exponential
distribution. It can be observed that increasing $k_{max}$ the
asymptotic linear part of $\sigma(\ep)$ is preceded by a longer and
longer plateau regime indicating a plastic response of the system. 
Figure \ref{fig:remaining} presents the remaining permanent
deformation $\ep_r$ of the bundle measured after unloading the system from the
maximum deformation $\ep_m$ of the loading process. One can observe
that  $\ep_r$ tends to the maximum value $k_{max}\left<\ep_{th}\right>$ as $\ep_m$ increases.
\begin{figure}
  \begin{center}
  \epsfig{bbllx=40,bblly=475,bburx=350,bbury=750,
file=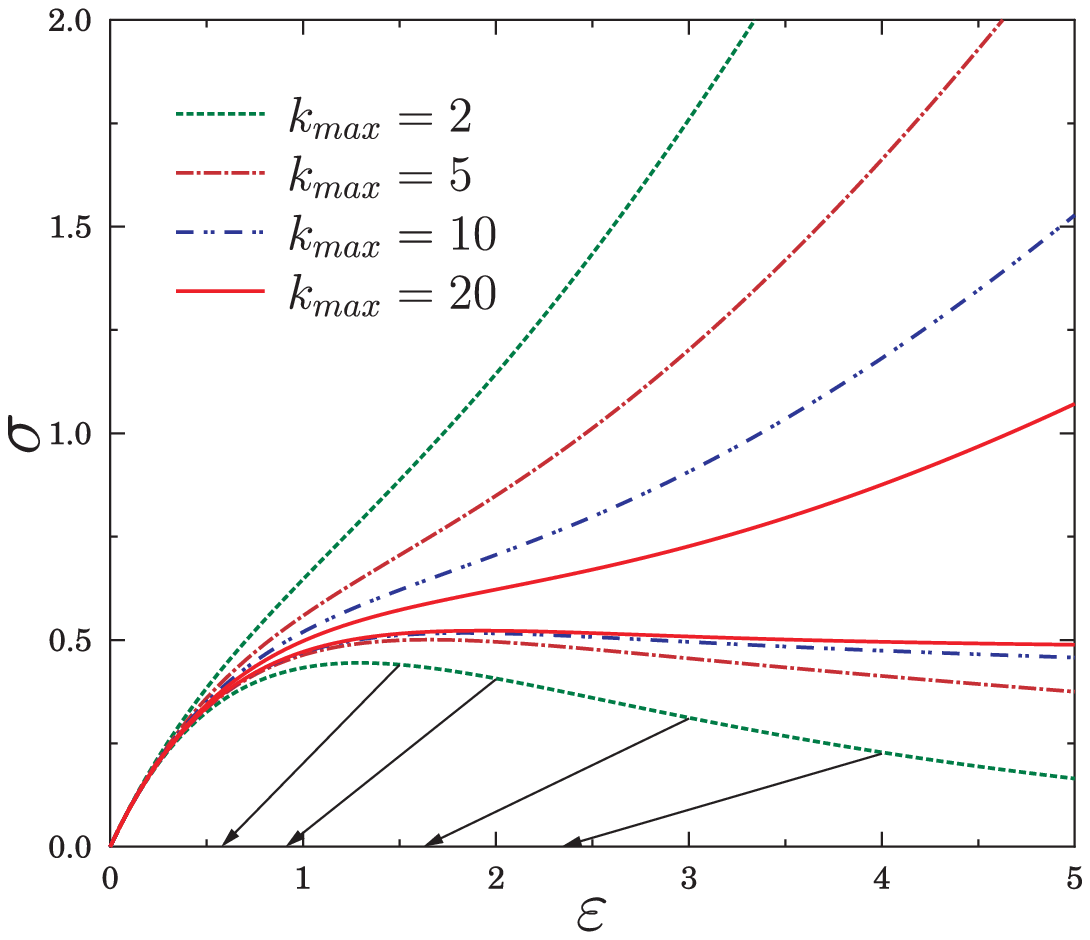,width=7.5cm}  
   \caption{({\it Color online}) Comparison of the constitutive curves
of the system with hardening (no failure) and
failure after $k_{max}$ sliding events. The arrows represent unloading
curves for $k_{max}=2$ with a decreasing Young modulus.} 
  \label{fig:constit_breaking}
  \end{center}
\end{figure}

In realistic situations it is typical that after a certain number of
sliding events the relaxed length of the subunits of the system cannot
be further extended, which then results in breaking. Such cases can be
captured by the model with the 
assumption that after a fiber has slided $k_{max}$ times it breaks
irreversibly, i.e.\ its stiffness is set to zero. The constitutive
equation of the sliding-breaking system can be obtained analytically
by skipping the last term of Eq.\ (\ref{eq:bundle_constit}) 
which has the consequence that for large deformations
$\ep\to\infty$ the stress tends to zero $\sigma(\ep)\to 0$ indicating that
the bundle gradually looses its load bearing capacity. Figure
\ref{fig:constit_breaking} presents a comparison of constitutive
curves Eq.\ (\ref{eq:bundle_constit}) with
hardening and failure after $k_{max}$ slidings. It is important to
emphasize that due to the breaking of fibers, the unloading Young
modulus of the system is a decreasing function of the maximum
deformation $\ep_m$ achieved before the load is released so that the
remaining deformation $\ep_r$ takes the form $\ep_r =
\ep_m-\sigma(\ep_m)/E[1-P(E\ep_m/k_{max})]$. In Fig.\
\ref{fig:constit_breaking} the arrows indicate unloading curves where
the decreasing unloading modulus can also be observed. Under stress
controlled loading the constitutive curve $\sigma(\varepsilon)$ can
only be realized up to the maximum, where macroscopic failure
occurs. The value of the maximum defines the fracture strength
$\sigma_c=\sigma(\varepsilon_c)$ of the sliding-breaking bundle, where
the critical deformation $\varepsilon_c$ can be determined numerically
from the equation
$1=\varepsilon_c[\sum_{k=1}^{k_{max}-1}(1/k)p(E\varepsilon_c/k)
-(k_{max}-1)/k_{max}p(E\varepsilon_c/k_{max})]$. Numerical analysis of
the above equation shows that both $\varepsilon_c$ and $\sigma_c$ are
increasing functions of $k_{max}$ converging to finite values.

In the case of annealed disorder, the fibers get a new threshold value
from the same probability distribution after each sliding event (see
also Fig.\ \ref{fig:single_fiber}$(b)$). The
new threshold may represent that the physical
properties of the new conformation attained by the fiber are different
from the previous ones including also the effect of thermal noise. The
macroscopic load $\sigma$ on the system 
can be obtained by summing the load
above subsets of fibers with different sliding numbers $k=0,1,\ldots ,
k_{max}$ as $\displaystyle{\sigma(\ep)=\sigma_0(\ep)+\sigma_1(\ep)+\cdots
+\sigma_{k_{max}}(\ep)}$. 
Due to the independence of subsequent thresholds the terms of the
above constitutive equation can be cast in the forms
$\displaystyle{\sigma_0(\varepsilon) = E\ep\left[1-P(E\ep) \right]}$, 
\beq{
\displaystyle{\sigma_1(\varepsilon) = \int\limits_{\ep_{th}^{min}}^{\ep}
p(\ep_1)d(\ep_1)E(\ep-\ep_1)(1-P(\ep-\ep_1)),}         
\label{eq:constit_5} 
}
\beqa{
&&\sigma_2(\varepsilon)=\int\limits_{\ep_{th}^{min}}^{\ep}
\int\limits_{\ep_{th}^{min}}^{\ep-\ep_1}d\ep_1d
\ep_2p(\ep_1)p(\ep_2)\\ 
&& \times \left[1-P(\ep-\ep_1-\ep_2)\right]E(\ep-\ep_1-\ep_2).      
\label{eq:constit_7}  \nonumber }
The general case of arbitrary $k<k_{max}$ reads as
\beqa{
&&\sigma_k(\varepsilon)=\int\limits_{\ep_{th}^{min}}^{\ep}
\int\limits_{\ep_{th}^{min}}^{\ep-\ep_1}\nonumber  
\cdots\int\limits_{\ep_{th}^{min}}^{\ep-\ep_1-\ep_2\cdots-\ep_k}
\prod_{l=1}^kd\ep_lp(\ep_l)
\\
&&
\times\left[1-P(\ep-\sum_{l=1}^k\ep_l)\right]E(\ep-\sum_{l=1}^k\ep_l),    
\label{eq:constit_8} } 
Finally, for $k=k_{max}$ we get
\beqa{
&& \sigma_{k_{max}}(\varepsilon) = \int\limits_{\ep_{th}^{min}}^{\ep}
\cdots\int\limits_{\ep_{th}^{min}}^{\ep-\ep_1-\cdots-\ep_{k_{max}}}\nonumber
\prod_{l=1}^{k_{max}}d\ep_lp(\ep_l) \\
&& \times E(\ep-\sum_{l=1}^{k_{max}}\ep_l), 
\label{eq:constit_9} }
Figure \ref{fig:const_exp} shows also examples of the constitutive
curve $\sigma(\ep)$ of the fiber bundle with annealed disorder. We
obtain qualitatively the same behavior with the 
same asymptote as with quenched disorder, however, the plastic plateau
becomes significantly longer. The result shows that the extension of
the plastic regime is determined by $\ep_r^{max}$, however, the
precise functional form of $\sigma(\ep)$ is sensitive to the type of
disorder.  

We presented a micro-mechanical model for systems which can extend
their length in a series of stick-slip events when subjected to an
external load. The model is an extension of fiber bundle models in
such a way that single fibers have stick-slip rheology characterized
by quenched or annealed threshold values of the local strain. We
showed analytically that varying its parameters the model provides a
broad spectrum of constitutive behaviors: for a large number of
sliding events a plastic regime develops which is then followed by
hardening (no breaking) and softening (fiber breaking). Unloading the
system a permanent deformation remains which is a monotonically
increasing function of the maximum deformation. If the fibers break
after $k_{max}$ sliding events, the unloading modulus goes to zero
with increasing deformation. 
The constitutive curves provided by the model have qualitative
agreement with the measured response of stick-slip systems
\cite{spider_vollrath_softmatter2006,fiber_bio_2007}, 
however, for a quantitative comparison further improvement is needed: 
allowing for an increase or decrease of the
Young modulus of single fibers after local sliding events would
make the model more realistic. Assembling bundles of fibers to form a
chain (serial coupling of fiber bundles) is also a promising
direction to reproduce the mechanical response of biological
tissues. Work in this direction is in progress.

F.\ Kun acknowledges the Bolyai Janos research fellowship of the Hungarian
Academy of Sciences.

\bibliography{/home/feri/statphys_fracture}

\end{document}